\documentclass[12pt,a4paper]{article}
\usepackage[english]{babel}
\usepackage{graphicx}

\newcommand{\alt}{\mathbin{\lower 3pt\hbox
   {$\rlap{\raise 5pt\hbox{$\char'074$}}\mathchar"7218$}}}
\newcommand{\agt}{\mathbin{\lower 3pt\hbox
   {$\rlap{\raise 5pt\hbox{$\char'076$}}\mathchar"7218$}}}

\begin{document}
\setcounter{footnote}{0}
\setcounter{equation}{0}
\setcounter{figure}{0}
\setcounter{table}{0}
\vspace*{5mm}

\begin{center}
{\large\bf Computer Model of a "Sense of Humour".\\
II. Realization in Neural Networks }

\vspace{4mm}
I. M. Suslov \\
P.N.Lebedev Physical Institute,\\
119334 Moscow, USSR\,\footnote{\,Present
address:

\noindent
P.L.Kapitza Institute for Physical Problems,
\\
119337 Moscow, Russia \\
E-mail: suslov@kapitza.ras.ru}
\vspace{4mm}
\end{center}

\begin{center}
\begin{minipage}{135mm}
{\large\bf Abstract } \\
The computer realization of a "sense of humour" requires the creation
of an algorithm for solving the "linguistic problem", i.e. the problem of
recognizing a continuous sequence of polysemantic images. Such an algorithm
can be realized in the Hopfield model of a neural network, if it is
suitably modified.
 \end{minipage} \end{center}
 \vspace{5mm}

 In [1] we analysed the general algorithm of information processing and showed
that on fulfillment of the natural requirements imposed by its
biological purpose such an algorithm will possess a "sense of
humour". The present paper proposes a possible realization of the
algorithm in a system of formal neurons.

\vspace{6mm}
\begin{center}
{\bf Description of the model }
\end{center}
\vspace{3mm}

 Following Hopfield [2] we shall consider that the state of the
 $i$-th neuron is described by the variable $V_{i}$ assuming two
values: $V_i=1$ (excited state) and $V_i = 0$ (state of rest).
The link of the neuron $i$ with the neuron $j$ is determined by
the parameter $T_{ij}$. The state of the system changes
with time $t$ according to the algorithm:
$$
V_i(t+\delta t)= \frac{1}{2}+\frac{1}{2} {\rm sign}
\left\{ \sum_{j} T_{ij} V_j(t)- U_i \right\}\,,
\eqno(1)
$$
where
$U_i$ is the excitation threshold of the $i$-th neuron and the
number $i$ is chosen randomly.

The proposed model of the nervous system is a modification of the trilayer
perceptrone [3] adapted for work in real time. It contains the following
elements (Fig.\,1).

{\it The associative memory ($A$\,--\,layer)} represents the
neural network which for simplicity we consider as described by
the Hopfield model: the neurons of the $A$\,--\,layer are linked
with each other with $T_{ij}=T_{ji}$, $ U_i=0$. Within the
$A$\,--\,layer evolution according to (1) leads from the
arbitrary initial state $\{V_i\}$ to one of the local minima of
energy
$$
E=-\sum_{ij}\, T_{ij}\, V_i\, V_j  \,,
\eqno(2)
$$
\begin{figure}
\centerline{\includegraphics[width=5.1 in]{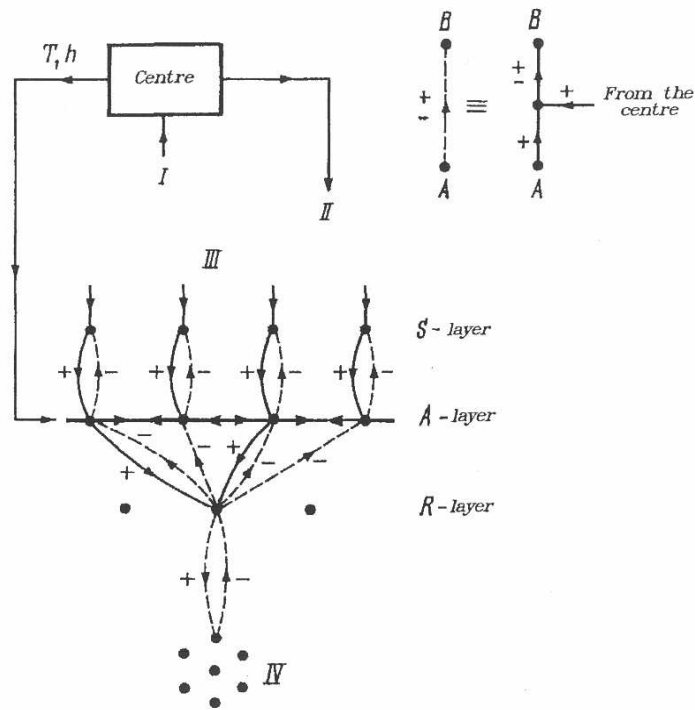}} \caption{
Proposed model of the nervous system is a modification of
the three\,--\,layer perceptrone [3]: filled circles are neurons;
solid lines are constantly acting links between neurons; broken
lines are links switched on the command from the centre; the
symbols $+$ and $-$ indicate the excitatory and inhibitory
character of the links; the $S$\,--\,layer is the sensory system;
the $A$\,--\,layer is the associative memory; the $R$\,--\,layer
is the reacting system or "consciousness". Roman numerals mark:
$(I)$ information from the $A$\,--\,layer, $(II)$ the control of
the links, $(III)$ the outer world, $(IV)$ the motor cortex.
Inset in the right upper corner shows possible realization of the
controllable link. } \label{fig1}
\end{figure}
which are identified with the images written in the memory
$\{V_{ij}^s\}$, $s = 1, 2,\ldots ,p$.
The recorded images determine the matrix of
the links $T_{ij}$ [2]:
$$
T_{ij} =\sum_{s=1}^{p}\, \mu_s\, (2 V_i^s-1)\,(2 V_j^s-1),
\qquad i\ne j \quad (\mu_s>0), \quad T_{ii}=0
\eqno(3)
$$

{\it The sensory system ($S$\,--\,layer)} receives signals from
the outside world (for example, from the retina of the eye). The
sensory neurons are not linked between themselves, but each
$S$\,--\,neuron is bound to one of the memory neurons:  the links
$S\to A$ are positive (exciting) and the back links $A\to S$ are
negative (inhibiting) (Fig.\,1).

\vspace*{2mm}

{\it The reacting system ($R$\,--\,layer)} consists of a set of
neurons each of which corresponds to one of the images recorded
in the memory: to the $s$-th neuron of the $R$\,--\,layer
converge the positive (exciting) links from those neurons $i$ of
the memory for which $V_{i}^{s}=1$  (we shall refer the last
neurons as a carrier of the image  $\{V_{i}^s\}$). The back
negative (inhibiting) links run from the $R$\,--\,neurons to the
memory neurons; the $R$\,--\,neurons are not linked between themselves
(Fig.\,1). The thresholds $U_i$ for the $R$\,--\,neurons are adjusted
in such a manner
that the excitation of the $s$-th neuron of the $R$\,--\,layer
occurs only when the configuration of the $A$\,--\,neurons is
sufficiently close to the image $\{V_{i}^s\}$.

 We consider that the image is realized by the biological
 individual only when there is excitation of the corresponding
$R$\,--\,neuron, i.e. the $R$\,--\,layer represents the consciousness
of the individual.

\vspace*{2mm}

{\it The centre} coordinates the work of the system by acting
according to the built-in program: it exercises control over the
macroscopic parameters of the system. The
concrete functions of the centre consist in the following.

(1) The centre has links with a small part of the memory neurons
evenly distributed in the $A$\,--\,layer which allows it to
register (a) the presence of excited
neurons in a certain portion of the memory and (b) the
stationarity or nonstationarity of this portion.

(2) The centre produces local changes of "temperature" in the
$A$\,--\,layer.  Since the temperature of the neural network is
determined by the noise level in it (which can be taken into
account by introducing the random force $f_{i}(t)$ into the
braces (1)) then the regulatable noise source should be at the
disposal of the centre.

(3) The centre produces local switching on the "magnetic field"
in the $A$\,--\,layer which corresponds to the addition
in  energy  (2)  of the term
$$
\sum_{i} h_i V_i
\eqno(4)
$$
(in (1) $h_i$ are added to the thresholds $U_j$). Switching on
the field is achieved with the aid of a "magnet" --- group of
neurons controlled from the centre, each of which is connected
with a certain region of the $A$\,--\,layer.

(4) The centre carries out the control of the links shown in
Fig.\,1 by a broken line. The simplest realization of the
controllable link $AB$ is possible with the aid of the
intermediate neuron $C$ (see Fig.\,1), the threshold of which is
so adjusted that it is excited only in the simultaneous presence
of the exciting signal from the neuron $A$ and from the centre.
In the presence of a signal from the centre the neuron $A$
excites or inhibits the neuron $B$ --- the link is switched on,
in the absence of a signal from the centre the neuron $A$ cannot
act upon the neuron $B$ --- the link is switched off. The command
for switching on and off is given not to the individual links but
simultaneously to their large groups.

(5) The centre gives the command for the learning of the plastic links.

\vspace{6mm}
\begin{center}
{\bf Recognition of a separate image}
\end{center}
\vspace{3mm}

 Recognition of the images occurs with the links $A\to S$ and $ R \to A$
switched on.  In the initial state of the system all the neurons are
not excited. Since the state of the $A$\,--\,layer with
$V_{i}\equiv 0$ is unstable  (see (1) at $U_i=0$), so
the presence of the stabilizing magnetic field is necessary.

 Let the stimulus $\tilde B=B+\delta B$  (i.e.  the "noisy" image $B$)
 is presented to the
 sensory system; this induces
 excitation of some of the $S$\,--\,neurons (Fig.\,2,a). Then the
\begin{figure}
\centerline{\includegraphics[width=5.1 in]{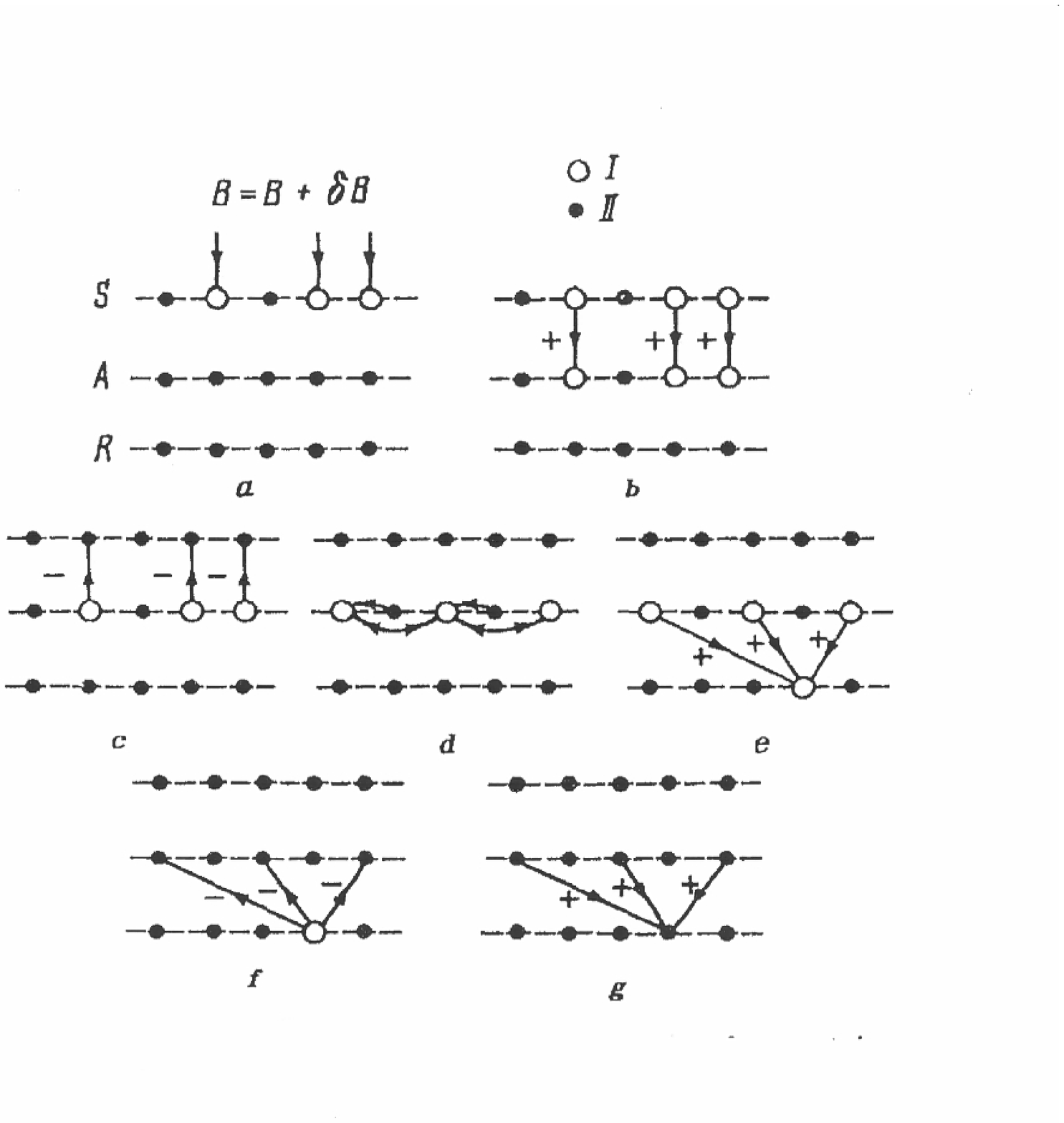}} \caption{
Sequential states of the system in the course of
recognition of a separate image; open circles $(I)$ are excited
neurons, dark circles $(II)$ --- not excited ones. For clarity,
only the links are shown along which excitation was transmitted
at the preceding moment of time. } \label{fig2}
\end{figure}
 centre switches out the magnetic field and excitation is
transmitted to the $A$\,--\,neurons (Fig.\,2,b) which, in turn,
quench the sensory neurons (Fig.\,2,c) (it is assumed that the
links $S\to A$ and $A\to S$ are suffiiciently strong).  Then in
the $A$\,--\,layer there is free evolution according to (1) which
ends in relaxation to the stable state corresponding to the image
$B$ (Fig.\,2,d); the neuron responsible for this image is excited
in the $R$\,--\,layer (Fig.\,2,e).  The back signal is send into
the memory, quenching the excited neurons (Fig.\,2,f) and the
image $B$  is deleted from the "consciousness" (Fig.\,2,g):  the
system returns to the initial state and is ready for the
perception of a new image.

\vspace{6mm}
\begin{center}
{\bf Learning}
\end{center}
\vspace{3mm}

The links between the $A$\,--\,neurons are plastic and change
according to the rule [2]
$$
\delta T_{ij}\sim (2 V_i-1)(2V_j-1) \delta t
\qquad  (i\ne j) \,,
\eqno(5)
$$
if the neurons $i$ and $j$ stay in the states $V_i$ and $V_j$ during the
time $\delta t$.  If in the initial state $T_{ij}\equiv 0$ then the
presentation to the system of $p$ configurations $\{V_{i}^s\}$,
$s= 1, 2, ...,p$  leads to the matrix of links (3). Since $T_{ij}$ may
have any sign, the neurons of the $A$\,--\,layer must have both
exciting and inhibiting synapses (specialization of the synapses
is known [4, pp.\,62--65] to have invariant character).

The links $A \to R$ in the initial state have a zero value and can be
learned only in one, positive direction
$$
\delta T_{ij} = \left\{
\begin{array}{cc} c \delta t\quad (c>0)\quad &
{\rm for } \quad V_1=V_j=1 \\
0\quad & ({\rm in\,\, other\,\, cases})
\qquad   \end{array} \right.\,,
\eqno(6)
$$
i.e. have only exciting synapses. The remaining links
($R\to A$, $S\to A$, $A\to S$) are not plastic
and have inborn character.

 The  learning of the system is similar to the learning of
  a child: a certain image $B$ is presented
 to it generating in the $S$\,--\,layer a certain configuration
 of excited neurons; then it is asked to "memorize $B$", which
excites one of the neurons of the $R$\,--\,layer appointed
to be responsible for the image $B$. Learning occurs with
the $A\to S$ and $R\to A$ links cut out (Fig.\,3), so that the
\begin{figure}
\centerline{\includegraphics[width=5.1 in]{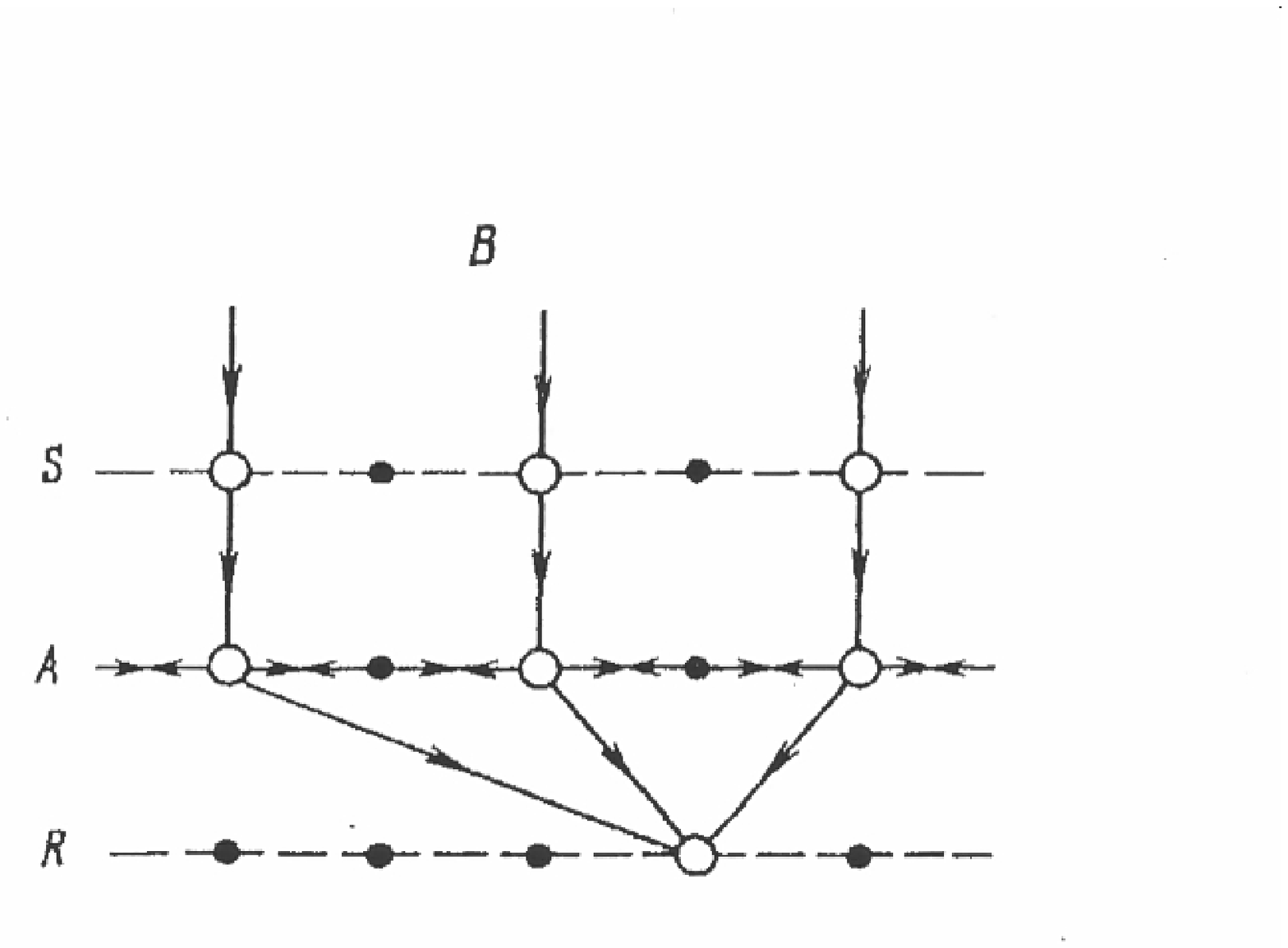}} \caption{
Learning occurs with the $S\to A$ and $R\to A$ links
switched off. } \label{fig3}
\end{figure}
configuration of the $S$\,--\,neurons is projected into the
$A$\,--\,layer and persists for a certain time: the links
$T_{ij}$ in the $A$\,--\,layer change according to (5) forming
the matrix (3), while the links $A\to R$ change according to (6)
ensuring the connection of the  $s$-th neuron of the
$R$\,--\,layer with a carrier of the image   $\{V_{i}^s\}$.

 In the resting state $V_i\equiv 0$ the system
  spends a considerable time and the learning according to
  (5) should result in  "ferromagnetic" interaction
  between the neurons ($T_{ij}>0$  for all $i,j$) and
  the single state $V_i\equiv 1$ being stable.  Therefore, we
suppose that there is no learning in the state
$V_i\equiv 0$, and the learning command is given only on
presentation of the image.

\vspace{6mm}
\begin{center}
{\bf Short-range action of the links and localization of images}
\end{center}
\vspace{3mm}

 In the usual Hopfield model [2] all the neurons are linked with
each other
and the carriers of images $\{V_{i}^s\}$ are spread out over the
whole neural net. In
reality the links $T_{ij}$ have a finite radius of action $\xi$: in
the human brain each neuron has $\sim 10^4$ synapses, while a complete
number of neurons $\sim 10^{11}$ [4, pp.\,31--33]. Experimental
indications on localization of the images also exist [3,
pp.\,64,\,65].

 We shall consider that each image $\{V_{i}^s\}$ is written
in a certain region $\Omega_s$ containing many neurons, but
small as compared
with the size of the whole neural network; so that
$V_{i}^s =0$  for $i$ not belonging
to $\Omega_s$ (the carrier $\{V_{i}^s\}$ is localized in $\Omega_s$)
and for $i\in \Omega_s$  the magnitudes $V_{i}^s$ with equal
probability assume the values 0 and 1.\,\footnote{\,To raise the
stability of the system relative to the
destruction of some of the neurons the region $\Omega_s$ can be
multiply connected.}
Since recognition of the images in
any event must be preceded by translational shift, rotation and
change of the scale (which can be achieved by a certain modification
of the Hopfield model [5]) then the assumption on the localization
of images does not have any serious consequences. The command for
switching off the magnetic field, change in temperature (see
below) and learning of plastic links is given only on
presentation of the image $\{V_{i}^s\}$  and only for neurons of
the region $\Omega_s$.

Taking all this into account, we accept that the learning rule
for the $A$\,--\,neurons instead of (5) has the form
$$
\delta T_{ij}\,\sim\, D_{ij}\,\delta_i^s  \delta_j^s
 (2 V_i-1)\,(2V_j-1)\, \delta t
\qquad  (i\ne j)
\eqno(7)
$$
when the image $\{V_{i}^s\}$ is presented;
here
$$
D_{ij} =\left\{ \begin{array}{cc} 1,\quad  &
{\rm for } \quad r_{ij}<\xi,\quad i\ne j \\
0,\quad & {\rm in\,\, other \,\,cases}\end{array} \right.\,,
\qquad
\delta_{i}^s =\left\{ \begin{array}{cc} 1,\quad   &
{\rm for } \quad i\in \Omega_s \\
0,\quad & {\rm for}\quad i\notin \Omega_s \end{array} \right.
\eqno(8)
$$
and $r_{ij}$ is the distance between the neurons $i$ and $j$.
The matrix of the links after writing $p$ images
instead of (3) assumes the form
$$
T_{ij} =D_{ij}\,\sum_{s=1}^{p} \,\delta_{i}^s\,\delta_{j}^s
 \mu_s (2 V_i^s-1)(2 V_j^s-1),
\qquad  \mu_s>0.
\eqno(9)
$$
To demonstrate stability of $\{V_{i}^s\}$ let us compose the
combination [2]
$$
H_i^{s}=\sum_{j} T_{ij} V_j^{s}=
\sum_{s'} \mu_{s'} \delta_{i}^{s'}  (2 V_i^{s'}-1)
\sum_{j\in \Omega_{ss'}}
D_{ij} V_{j}^{s}  (2 V_j^{s'}-1),
\eqno(10)
$$
where $\Omega_{ss'}$ is the intersection of $\Omega_s$ and
$\Omega_{s'}$. By virtue of
the randomness of $V_{i}^s$ and the large number of terms in
the sum for $j$ the latter  is close to its mean:
$$
H_i^{s}=\frac{1}{2}\mu_{s} \delta_{i}^{s}
\sum_{j\in \Omega_{s}} D_{ij} (2 V_i^{s}-1)
\eqno(11)
$$
For $i\in \Omega_{s}$ the sum over $j$ is positive and
$\delta_i^{s}=1$ so that the
configuration of $\{V_{i}^s\}$ is stable by virtue of
the algorithm (1); for $i\notin \Omega_{s}$
the state $V_i=0$ is maintained by the magnetic field.

Because of the localization of the images it suffices for the
$s$-th neuron of the $R$ \,--\,layer to have links only with
the neurons of the  region $\Omega_s$ of the $A$\,--\,layer.

\vspace{6mm}
\begin{center}
{\bf Recognition of the ambiguous image}
\end{center}
\vspace{3mm}

Above it was assumed that the stimulus presented,
$\tilde B=B+\delta B$, is close to image $B$ contained in
the memory so that the initial state $\tilde B$ always
relaxes to the final state $B$, i.e. the interpretation of
the image $\tilde B$ is
clearcut. In terms of energy this means that the state $\tilde B$
lies in a potential well, whose  minimum corresponds
to the image $B$ and evolution occurs at zero temperature.

 Now let us consider the recognition of an ambiguous image. We have in mind
the modelling of the following situation:
the word $B$ is shown to the person  and it is explained to him
that this word may assume several meanings: in the
memory of the person the images $B+b_1$, $B+b_2,\ldots$ are
fixed where $b_1$, $b_2,\ldots$ are the elements of the given
explanation; if now the word $B$ is presented for recognition its
interpretation will be ambiguous leading to one of the results
$B+b_i$. For modelling it suffices to assume that recognition
begins at the temperature $T$ exceeding the potential barriers between
images $B+b_1$, $B+b_2,\ldots$ and then the temperature decreases
and the system relaxes to
one of the local minima corresponding to the images $B+b_i$.
 Hereafter, we shall consider that decrease of temperature from
the initial value $T_0$ occurs adiabatically so that the system with
overwhelming probability relaxes to the deepest of the local minima.

\vspace{6mm}
\begin{center}
{\bf Simultaneous recognition of several images}
\end{center} \vspace{3mm}

Suppose that after presentation of the stimulus $A$
the system relax to one  of the images $A_1$, $A_2,\ldots$ and
after  presentation of the stimulus $B$ to one of the
images $B_1$, $B_2,\ldots$ What will happen, if the stimuli $A$
and $B$ are presented simultaneously?

 The process of  recognition begins with switching out the
 magnetic field in the portion of the $A$\,--\,layer, where the
 image carrier is localized. Let presentation of the images $A$
and $B$ requires switching out the field in the regions
$\Omega_A$ and $\Omega_B$ of the associative layer. Let us assume
for simplicity that the regions $\Omega_A$ and $\Omega_B$ do not
overlap (the qualitative picture persists in the general case);
then $V_i=V_i^A+V_i^B$ where $V_i^A$ and $V_i^B$ differ from zero
respectively for the neurons $i$ lying in the regions $\Omega_A$
and $\Omega_B$. The energy (2) assumes the form
$$
E\{V_i^A + V_i^B\}=-\sum_{ij} T_{ij} V_i^A V_j^A
-\sum_{ij} T_{ij} V_i^B V_j^B
-2\sum_{ij} T_{ij} V_i^A V_j^B
\eqno(12)
$$
If the stimuli $A$ and $B$ are presented separately,
then correspondingly $V_i^B\equiv 0$ or $V_i^A\equiv 0$, so that
only the first or second term remains
in the right part of (12).  The third term in (12) shows that
with the simultaneous presentation of the stymuli $A$ and $B$,
they exert on each other an effect equivalent to the presence of
a magnetic field. The result of such interaction is particularly
clear if the configurations $A_1$, $A_2,\ldots$ and
respectively $B_1$, $B_2,\ldots$  are almost degenerate. If the
links $T_{ij}$ between the regions $\Omega_A$ and $\Omega_B$ are
absent then the equilibrium states of the system have the form
$(A_s,B_{s'})$ and are almost degenerate.  Inclusion of the weak
links $T_{ij}$ between the regions $\Omega_A$ and $\Omega_B$
does not change significantly  the equilibrium configurations,
but changes the relative  position on the energy scale.
If on separate recognition of the images $A$ and $B$
certain configurations (e.g. $A_1$ and $B_2$ in Fig.\,4)
\begin{figure}
\centerline{\includegraphics[width=5.1 in]{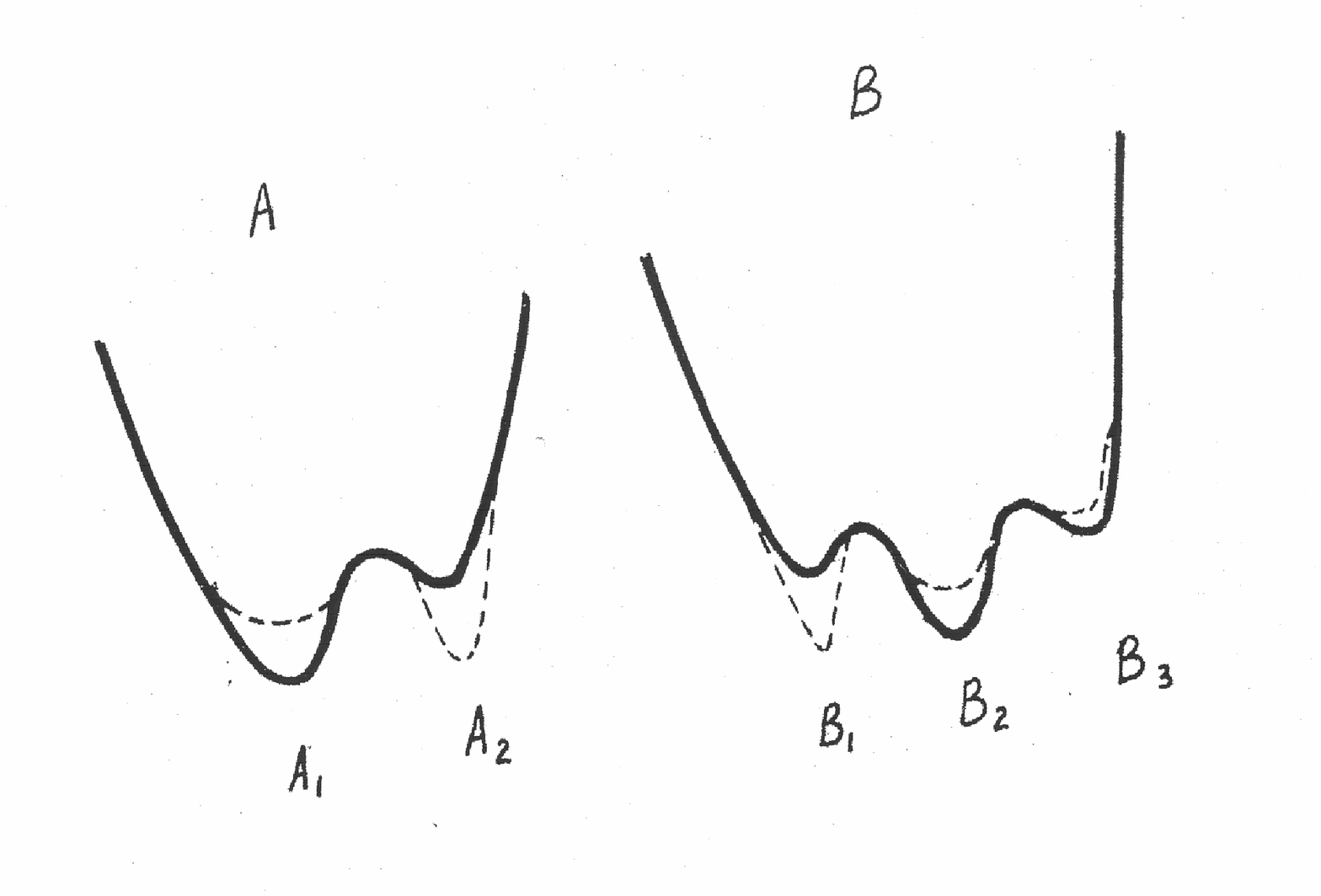}} \caption{
Potential relief for recognition of the images
 $A$ and $B$ separately (continuous curve) and for simultaneous
recognition (broken line). } \label{fig4}
\end{figure}
are energetically advantageous, then on their simultaneous
recognition  other configurations may prove more
advantageous (e.g. $A_2$ and $B_1$). It may be
visually imagined that the interaction change the potential
relief where $A$ and $B$ relax (Fig.\,4). As a result, the
interpretation of the polysemantic image proves to depend on the
context.

In the preceding paper [1], the existence of an algorithm was
postulated for the recognition of $N$ simultaneously presented
polysemantic images; now it is clear that such an
algorithm can be naturally realized in the Hopfield model in the
following conditions:  (a) recognition begins at finite
temperature $T_0$ which then adiabaticatty diminishes; (b) the
potential barriers between the meanings of the polysemantic image
are less than $T_0$; (c) the potential bariers between the
different polysemantic images are considerably greater than
$T_0$; and (d) the interaction between the images is
sufficiently weak.

\vspace{6mm}
\begin{center}
{\bf Recognition of a continuous sequence of images}
\end{center}
\vspace{3mm}

Above we assumed that all the images $\{V_{i}^s\}$  have localized
carriers.  Due to short-range nature of  links $T_{ij}$, the
associatively-related images should have closely positioned or
 overlapping carriers, while the uncorrelated images
have the carriers localized in the remote parts of the
memory.  Therefore, presentation of a continuous
sequence of  stimuli $A_1$, $A_2,\ldots$  give rise to
sequential excitation of the regions
$\Omega_1$, $\Omega_2,\ldots$  in the associative layer, which
looks like "diffusion" (Fig.\,5): the nearest in succession
\begin{figure}
\centerline{\includegraphics[width=5.1 in]{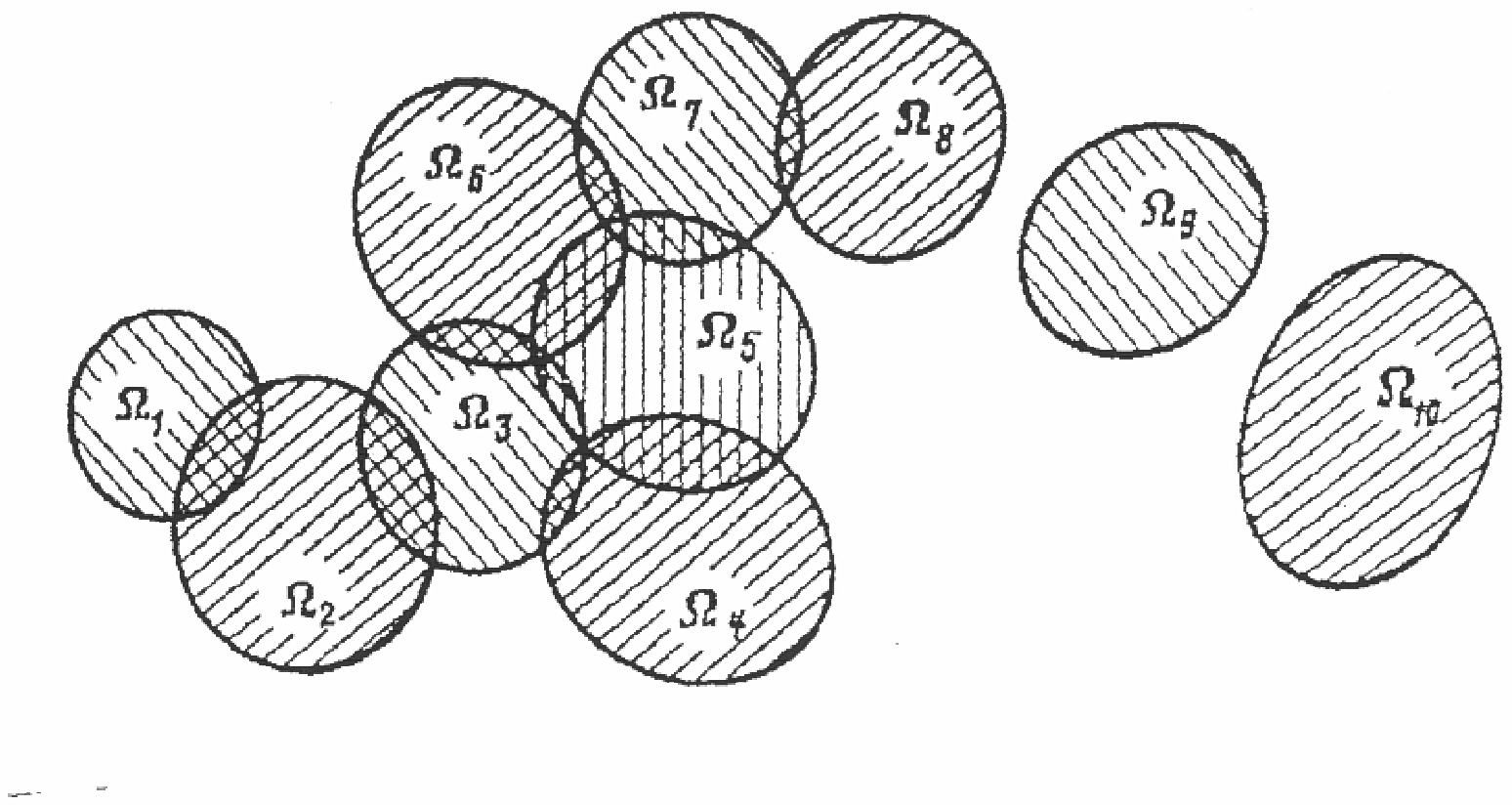}} \caption{
Recognition of a continuous sequence of images
$A_1$, $A_2,\ldots$: in the associative layer there is sequential
 excitation of the regions  $\Omega_1$, $\Omega_2,\ldots$,
  which looks fike "diffusion". } \label{fig5}
\end{figure}
images are correlated and their carriers form
conglomerations, whereas to the remote images in sequence
 correspond carriers remote in space, as a result of weakening of
 the correlations.  It gives the possibility to establish   the
 correspondence of the time of appearance of the image  with the
 position of its carrier in space, which greatly simplifies the work
 of the  centre.  After fixing the appearance of the excited neurons
 in a  certain portion of the $A$\,--\,layer, the centre raises the
 temperature of this portion to the value $T_0$, and after the
 characteristic time $\tau_0$ begins its adiabatic decreasing.
 This stimulates  establishment of the steady state, the
 bringing of the corresponding images into the $R$\,--\,layer and
 the zeroing of the corresponded portion  of the memory
 (Fig.\,2,d--g). Thereby, the trajectory of the  "diffusion"
 movement (Fig.\,5)  is "wiped off" after a certain time
 so that at each moment only its finite  segment exists in the
 memory; recognition of a continuous sequence of
 images is thereby reduced to the simultaneous recognition of a
 finite number of images (see above).

\vspace{6mm}
\begin{center}
{\bf Time delays and humorous effect}
\end{center}
\vspace{3mm}

 Let at the moment of time $t=0$ the neurons are excited in a
 certain portion of memory (Fig.\,2,b), at the moment $\tau_0$
the corresponding image is   brought into consciousness
(Fig.\,2,e) and at the moment $\tau_1$  the portion of  memory
 under consideration is
zeroed (Fig.\,2,f).  The delay $\tau_0$ corresponds to the
 interval $AC$ and the delay $\tau_1$  to the interval $AB$ in
 Fig.\,2 of the paper [1]; the latter is related to the fact that
 the possibility of reinterpreting the image persists until  the
corresponding portion of memory is nullified.  Obviously,
$\tau_1\ge \tau_0$:  this result was derived in [1] from
requirements of the optimal processing of the algorithm,
but now it holds due to the constructive features of the model.

The delay $\tau_0$ is determined by the rate of decreasing of
temperature (see  above), while the delay $\tau_1$ by the moment
of activation (from the  centre) of the $R\to A$ links. The
 optimal choice of the parameters $\tau_0$ and $\tau_1$ is
 determined by different principles:  the parameter $\tau_0$
corresponds to the delay from the moment information is received
by the brain till its appearance in consciousness and is upwardly
 limited by the  value $\tau_{max}$ \cite{1}, while the parameter
 $\tau_1$ regulates the loading of memory\,\footnote{\,Unlike
 the general case [1], in the present model there is no special
 operative memory.}, i.e.  the fraction of excited neurons in
 the $A$\,--\,layer (Fig.\,5).  This fraction should not be too
 small for the operative possibitities of the  brain to be used
 in full, and not too large for  interference of the images
 arriving at different times not to appear.

Let at the moment $t=0$ the image $A$ enters the memory; evolution in the
 corresponding potential relief (Fig.\,4, continuous curve) leads
 at the moment $t =\tau_0$ to stabilization of the memory
neurons in the configuration $A_1$ and excitation of the corresponding
 $R$\,--\,neuron. Let in the interval between $\tau_0$ and $\tau_1$
new image $B$ enters the memory and changes the potential relief
for $A$ (Fig.\,4, broken line). If the temperature at this
moment is sufficient to overcome the barrier, the
 system begins to relax to the configuration $A_2$ (in fact,
leaving of the state $A_1$ is possible even  at $T=0$ since
 the minimum corresponding to $A_1$ may disappear).  Such
 breaking of stationarity in a certain memory region after
 exciting of the corresponding  $R$\,--\,neuron  is considered as
 a sign of the humorous effect [1].  The image $A_1$ (or, in the
 general case,  a version consisting of several images) is
 realized as "false" and should be immediately deleted from
 consciousness; however, it cannot be done in the course of the
 usual routine (Fig.\,2,e,g) because of the need to obtain a new
 steady configuration $A_2$.

\vspace{6mm}
\begin{center}
{\bf  Mechanism of laughter}
\end{center}
\vspace{3mm}

 The "emergency" deletion of the false version from
 the $R$\,--\,layer is achieved
by activation of the links between the $R$\,--\,layer and
 the motor cortex (Fig.\,1); excitation of the $R$\,--\,neurons
 is transmitted to   the motoneurons and produce the
 contraction of certain muscles, i.e. laughter.

In fact we return  to the old idea by G. Spencer [6]
that the humorous effect is accompanied by release from the
mental process of nervous energy which is directed at the
muscular reaction.  This idea was supported by Darwin [7] and
Freud [8] but was criticized by later investigators [9] in view
 of the difficulty to introduce the concept of "nervous
energy". In fact, the definition of energy for neural networks
 may be given only under condition $T_{ij}=T_{ji}$  [2],
 which is  not  very realistic and does not hold for the
 considered model as a whole; so the concept of "nervous energy"
should not be taken seriously.  Nevertheless, the
qualitative picture agrees with Spencer's hypothesis: it looks
as if the energy of excitation is expelled
from the $R$\,--\,neurons into the motor cortex.

 The release of nervous energy in the presence of the
humorous effect was
validated by Spencer using the concept of "descending incongruity" ---
transition from a high to low style, i.e. from the state with
rich associations
to the state with poor associations. Such an interpretation
of the humorous
effect is surely limited and cannot lay claim to universality. In
the proposed scheme the "release of nervous energy" (in the
conditional sense indicated above) is related with the need to
delete the false version from consciousness, which
requires "zeroing" of a certain portion of the $R$\,--\,layer
(i.e. transition of the excited neurons to the nonexcited state).

 Since laughter is interpreted as an unconditioned reflex to the humorous
effect, the known cases of "ousting" of laughter by secondary emotions
require
an explanation. Laughter may be ousted by the emotions of indignation (an
indecent anecdote is told to a puritan),
fear (a bush suddenly turns ouf to be a bear),
compassion (a man in front of you
slips on a water melon rind and badly hurts himself), shame (you slipped on a
water melon rind) and so on. Within the Spencer hypothesis [6] all these
instances are explained by the fact that the "released nervous energy" is
directed not to the motoneurons but to other parts of the nervous system and
 goes on the formation of a secondary emotion
 (the $R$\,--\,layer is connected
not to the motor cortex but to the limbic system). The same ideas
[8] are used to explain the known fact that a joke produces the
greatest effect if it is told extremely laconically: laconicity
reduces the probability of formation of secondary associations
liable to absorb the "nervous energy".

 Casting the excitation of the neurons into different
portions of the motor
cortex, a man can regulate the level of the muscular reaction:
this can explain  its dependence on mood, the
psychological setting, the presence of a laughing enviroment [10]
and so on.
\newpage

\vspace{6mm}
\begin{center}
{\bf Conclusion}
\end{center}

The realization of a sense of humour requires a quite intricately
organized system. We would emphasize, however,
that this complex organization is
entirely governed by the task of treating a continuous sequence of polysemantic
images; the existence of the humorous effect is a secondary consequence.
It is well-known [4, pp.\,219--241], that different parts of the
brain have their own specialization; the proposed model may lay
claim to be a description of only the region of the brain where
the linguistic functions are concentrated  (so-called Broca
and Wernicke zones); other portions of the brain may have a
different organization.

The author is grateful to D. S. Chernavskii for fruitful discussions.

\end{document}